\documentclass[%
 reprint,
 amsmath,amssymb,
 aps,
]{revtex4-2}

\usepackage{graphicx}
\usepackage{dcolumn}
\usepackage{bm}
\usepackage[mathlines]{lineno}

\usepackage{color}

\newcommand{\revise}[1]{\textcolor{black}{#1}}

\begin{document}

\preprint{APS/123-QED}

\title{Observation of Galactic center in sub-MeV gamma-ray band with electron-tracking Compton camera}

\author{Tomonori Ikeda}
\email{tomonori.ikeda@aist.go.jp}
\affiliation{
National Institute of Advanced Industrial Science and Technology (AIST), National Metrology Institute of Japan (NMIJ),  1-1-1 Central 3, Umezono, Tsukuba, Ibaraki, 305-8563, Japan
}

\author{Toru Tanimori}
\affiliation{
 Department of Physics, Kitasato University,
1 Chome-15-1 Kitazato, Minami Ward, Sagamihara, Kanagawa, 252-0329, Japan
}
\affiliation{Graduate School of Science, Kyoto University Kitashirakawa Oiwakecho, Sakyo, Kyoto, Kyoto, 606-8502, Japan}

\author{Atsushi Takada}
\author{Taito Takemura}
\author{Kei Yoshikawa}
\author{Yuta Nakamura}
\author{Ken Onozaka}
\author{Mitsuru Abe}
\affiliation{Graduate School of Science, Kyoto University Kitashirakawa Oiwakecho, Sakyo, Kyoto, Kyoto, 606-8502, Japan}

\author{Yoshitaka Mizumura}
\affiliation{Institute of Space and Astronautical Science, Japan Aerospace Exploration Agency Yoshinodai 3-1-1, Chuou, Sagamihara, Kanagawa, 252-5210, Japan}


\date{\today}

\begin{abstract}
We report the direct detection of gamma-ray emission from the Galactic center in the 150--600~keV band using the electron-tracking Compton camera (ETCC), which has a wide field of view of 3.1~sr. This represents the first application of this linear, imaging-spectroscopy method to observations of the Galactic center.
Measurements in a one-day flight over Australia yielded significant gamma-ray detection in the light curve and revealed a $7.9\sigma$ excess over the background in the image map from the Galactic center region. 
These results, obtained through a simple and unambiguous analysis, demonstrate the high reliability and sensitivity of the ETCC and establish its potential for future high-precision MeV gamma-ray observations.
The measured intensity and spatial distribution were tested against three emission models: a single point-like source, a multi-component structure, and a symmetric two-dimensional Gaussian. All three were found to be statistically consistent with the data. 
\revise{The positronium-related flux provided by the multi-component model is $(3.2~\pm~1.4)~\times~10^{-2}$~photons~cm$^{-2}$s$^{-1}$, consistent with the value reported by INTEGRAL within $1\sigma$.}
\revise{These results establish the potential of the ETCC for future high-precision MeV gamma-ray surveys.}

\end{abstract}

\maketitle


\section{\label{sec:intro}Introduction}
The Galactic center and ridge are prominent sources of continuum hard X-ray and gamma-ray emission.
Studies of diffuse gamma-ray emission from the Milky Way, such as those by HEAO-3~\cite{Mahoney_1994}, COMPTEL onboard CGRO~\cite{comptel_1993}, and SPI onboard INTEGRAL~\cite{Bouchet_2011} have revealed a complex structure in the Galactic emission in the MeV band, attributed to a mixture of bremsstrahlung, inverse Compton scattering, and nuclear line emission. In particular, the detection of the 511~keV positron annihilation line by instruments like GRIS~\cite{Gehrels_1991}, OSSE~\cite{Purcell_1997}, and INTEGRAL~\cite{INTEGRAL_2003}, has provided strong evidence for positron production and annihilation in the interstellar medium. These observations have established that the Galactic center is the brightest source of 511 keV line emission, which suggests a concentrated source of positrons but leaves their origin uncertain.

Several candidates for Galactic positrons have been proposed, including astrophysical sources producing $\beta^{+}$ decay nuclei such as massive stars~\cite{Diehl2006}, core-collapse supernovae~\cite{Tsygankov_2016}, and Type $\rm{I}$a supernovae~\cite{Diehl_2015}; and compact objects producing positrons by photon-photon pair interactions such as neutron stars and black holes~\cite{Siegert_2016_2}.
However, the observed distribution of the 511 keV emission does not match that of any known astrophysical population and shows little correlation with other wavelengths.
This discrepancy has led to development of alternative hypotheses, including exotic physics scenarios. Among them, dark matter annihilation or decay remains a compelling possibility. Recent studies have proposed that low-mass primordial black holes (PBHs) could serve as a source of Galactic positrons through Hawking radiation~\cite{Keith_2021,Siegert_pbh_2022}. 

The latest measurements from the INTEGRAL observatory have shown that the 511~keV line emission from the annihilation of positrons has a multiple components: a point-like source from the Galactic center, a narrow bulge, a broad bulge, and a low surface-brightness disk~\cite{Skinner_2014, Siegert_2016}. 
The Compton Spectrometer and Imager (COSI)~\cite{tomsick_2023} observed the 511~keV line emission and the bulge component~\cite{Kierans_2020}. 
The observed extent was 2--3 times larger than the INTEGRAL result, using predefined emission templates~\cite{Siegert_2020}.
This broader distribution is consistent with earlier measurements by WIND/TGRS~\cite{Harris_1998}.
One critical limitation of coded aperture mask instruments, such as those on INTEGRAL, is their poor discrimination of isotropic or halo-like emission from the instrumental background.
Because the coded mask responds similarly to uniform sky emission and background, it becomes difficult to disentangle the two.
While conventional Compton telescopes offer improved imaging capabilities, their performance is degraded by the lack of directional information from the recoil electron, resulting in a limited point spread function (PSF).

The electron-tracking Compton camera (ETCC)~\cite{Tanimori_2015,TANIMORI2004263} records complete Compton kinematics information, enabling both a linear response and a well-defined PSF without requiring non-linear imaging algorithms such as the maximum entropy method~\cite{Strong1995} and other imaging techniques~\cite{Knodlseder_1999, Kierans2022, IKEDA201446, Wilderman_1998}.
In 2018, we conducted a balloon experiment using the ETCC, referred to as SMILE-2+, which successfully observed the Crab Nebula with a detection significance of 4.0$\sigma$~\cite{2022Takada}. 
Furthermore, the field of view (FOV) of the ETCC included the Galactic center region during the floating flight.


In this paper, we report the direct measurement of the gamma-ray intensity in the Galactic center region obtained with the ETCC, along with the results of statistical tests evaluating the spatial distribution of the emission in the energy range from 150~keV to 600~keV. The linearity of the ETCC response enabled background-subtracted skymap generation and imaging-spectroscopy analysis. This represents a fundamental distinction from previous MeV gamma-ray observations, which typically extract signals through template fitting.

\section{Methods}\label{sec:methods}
The ETCC consists of a micro-pattern gaseous time projection chamber ($\mu$TPC), which acts as a Compton-scattering target, and pixel scintillator arrays (PSAs), which absorb and measure the scattered gamma-rays. The $\mu$TPC has an active volume of 30~$\times$~30~$\times$~30~cm$^{3}$ and 108 PSAs made of GSO (Gd$_{2}$SiO$_{5}$:Ce) crystals are arranged around it, each with 8~$\times$~8 pixels. Further details of the on-board instrumentation are described in Ref.~\cite{2022Takada}.

\revise{The ETCC imaging performance is quantified by the half-power radius (HPR) of the PSF, defined as the angular radius within which 50\% of the reconstructed gamma-ray events from a point source are contained in the forward imaging~\cite{Tanimori2017, Tanimori_2015, Bernard2022}.
This definition is based on geometrical optics and is consistent with those used in optical, X-ray, and GeV gamma-ray telescopes. 
The PSF is defined for each incident gamma-ray, and the image brightness is determined accordingly. As a result, the reconstructed image preserves high linearity and quantitative reliability. The ETCC is the first instrument to establish a rigorous PSF definition for gamma rays in the MeV band and to enable image analyses based on linear imaging principles widely used in  astronomy.}

\revise{In contrast, the angular resolution measure (ARM), commonly used in Compton telescopes, characterizes the width of the Compton scattering angle distribution.
The ARM does not represent the width of the PSF in forward imaging and therefore differs fundamentally from the geometrical optics-based definition of angular resolution.
Because the scatter-plane deviation (SPD)~\cite{Tanimori_2015}  cannot be properly measured in conventional Compton cameras, a PSF cannot be defined in the same sense as in optical systems. Consequently, an image intensity cannot be uniquely assigned to each spatial direction, and linear image analysis methods analogous to those used in optics are not directly applicable. Instead, imaging performance must be evaluated statistically, requiring modeling of large data sets, source distributions, and background noise.
Accordingly, the angular resolution that characterizes the imaging capability of the ETCC is fundamentally different from the ARM.}

To improve the accuracy of the recoil electron direction, we applied a deep learning method based on convolutional neural networks~\cite{Ikeda2021}. The effective area and the HPR at 511~keV are estimated to be 0.59~cm$^2$ and 20~degrees, respectively.
\revise{The energy resolution, defined as the full width at half maximum (FWHM), is 14\% at 511~keV. 
Detailed measurements of the energy resolution as a function of energy are described in Ref.~\cite{2022Takada}.}

The SMILE-2+ balloon flight was launched from Alice Springs, Australia, on 2018-04-07T06:24 (format $YYYY$-$MM$-$DD$T$hh$:$mm$ where $YYYY$ is year, $MM$ is month, $DD$ is day of month, $hh$ is hour, and $mm$ is minute) Australian Central Standard Time (ACST).
The ETCC captured the Galactic center within the FOV of a zenith angle below 60$^{\circ}$ on 2018-04-08T01:00 ACST, and it remained observable for approximately 5 hours.

To suppress contamination from atmospheric gamma-rays and cosmic rays, we applied the event selection criteria described in Ref.~\cite{2022Takada}.
Additionally, we restricted the FOV to zenith angles below 60$^\circ$ to reduce the atmospheric gamma-ray background.
The details of the efficiency of the gamma-ray selection and background contribution were discussed in Ref.~\cite{2023Ikeda}. 

We binned the data both spatially and energetically. The skymap was divided into 12 pixels in Galactic coordinates, each covering approximately 33$^\circ$ corresponding to about 1~sr, which is comparable to the PSF of the ETCC.
The energy range from 150~keV to 600~keV was divided into four bins: 150--250~keV, 250--350~keV, 350--450~keV, and 450--600~keV.

The gamma-ray data collected by the ETCC consist of gamma-rays from point sources and the Galactic diffuse emission in the FOV, convolved with the detector response. 
In addition, the extragalactic gamma-ray background, atmospheric gamma-rays, cosmic rays, and accidental events contribute as background.
The expected count $D^{E',p'}$  in the detected energy bin $E'$ and sky pixel $p'$ is expressed as:
\begin{eqnarray} \label{eq:conv}
    D^{E',p'} =&&\sum_{E} \sum_{t} \sum_{\Omega} R^{E',p'}_{E}(t) (S^{E}(l,b) + P^{E}(l,b)) \Delta\Omega \Delta t \nonumber \\ 
    &&+ \sum_{t} A^{E'}B^{E',p'}(t) \Delta t, 
\end{eqnarray}
where $R^{E',p'}_{E}(t)$ is the time-dependent response function of the ETCC, including the atmospheric response,  $S^E(l,b)$ and $P^E(l,b)$ are the sky distributions of the Galactic diffuse emission and point sources, respectively, and $B^{E',p'}(t)$ is the background model scaled by the normalization coefficient $A^{E'}$. 
The index $E$ is the bin number of the incident gamma-ray energy.
We divided the incident gamma-ray energy into the same four bins as the energy interval of the observed energy.


\begin{figure}
\includegraphics[width=8.5cm]{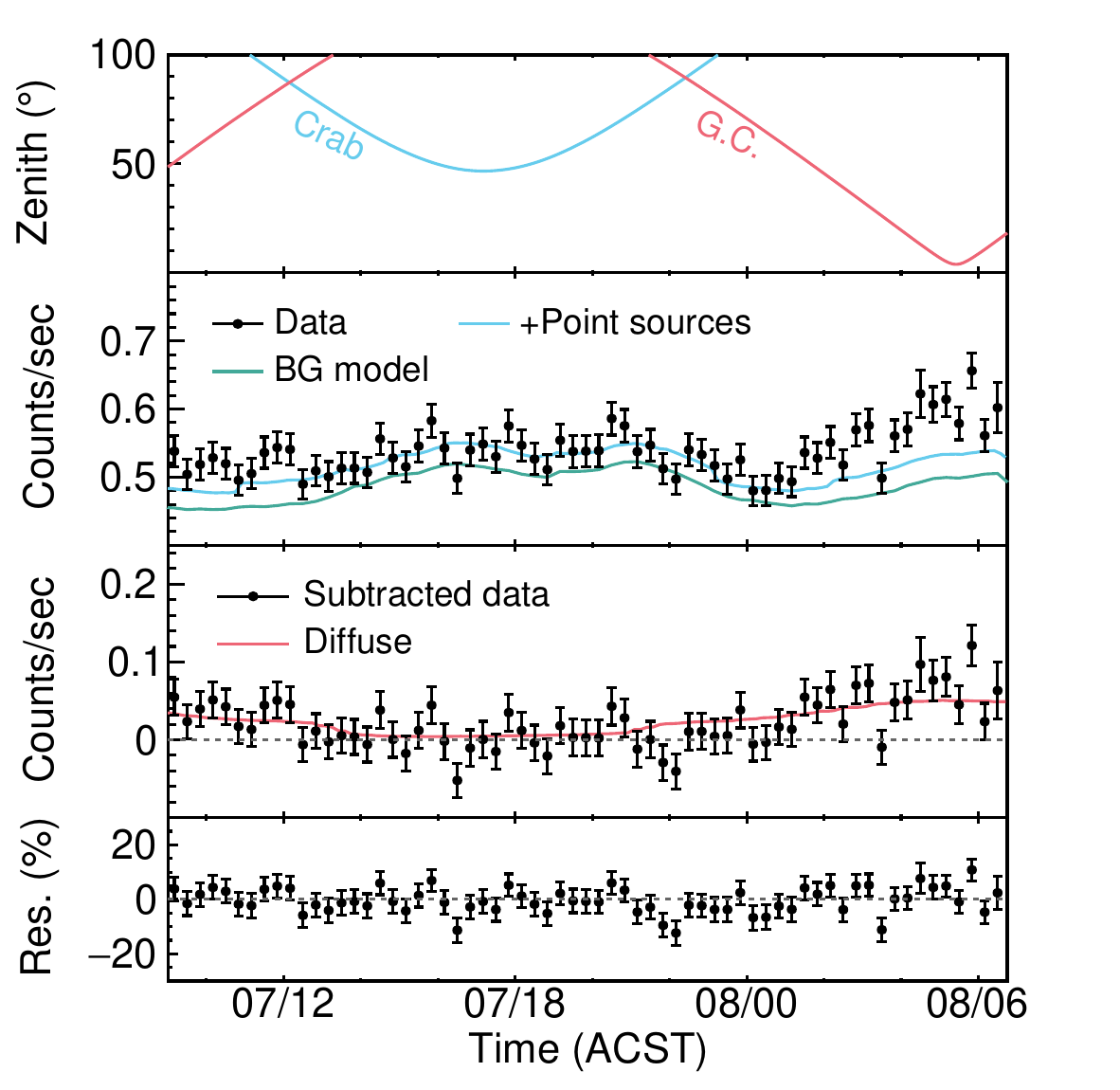}
\caption{\label{fig:lightcurve} 
Top panel: Zenith angles of the Crab Nebula (blue) and the Galactic center (red) as a function of time. 
Second panel: Observed gamma-ray count rate (black points with error bars). The green line shows the estimated background model.  The blue line adds the contribution from known point sources.
Third panel: Count rate after subtracting the background and point-source contributions , overlaid with the predicted diffuse emission of multi-component model (red).
Bottom panel: Residuals between the observed data and the total model (including diffuse emission), expressed as a percentage.}
\end{figure}

The background has mainly three contributions: the atmospheric gamma-rays, the cosmic ray events, and the accidental events~\cite{2023Ikeda}. 
\revise{The ETCC requires a coincidence trigger between the PSAs and the $\mu$TPC. The coincidence window is 9.5~$\mu$s. Therefore, isolated triggers in each detector can accidentally generate an ETCC trigger within this time window, even though the events are not physically correlated. Such accidental events are therefore defined as random coincidences between physically uncorrelated triggers.}

The energy range of 150--600~keV is dominated by atmospheric gamma-rays. 
Each event rate depends on the atmospheric depth and cut-off rigidity. 
Hence, we generated a background model $B^{E',p'}(t)$, that depends on the atmospheric depth and cut-off rigidity, following the procedure in Ref.~\cite{2023Ikeda} with PARMA~\cite{PARMA} and Geant4~\cite{Geant4}.
However, a slight discrepancy was observed between the total count rate in the flight data and that predicted by the simulated background model. Therefore, the normalization coefficient $A^{E'}$ was determined by fitting the light curve.
The fitting window was defined as the time interval from 2018-04-07T14:00 ACST to 2018-04-07T21:00 ACST, during which the Galactic center was outside the FOV.
Meanwhile, bright point sources including the Crab Nebula entered the FOV. Their contributions to the total count rate were estimated using the Swift Burst Alert Telescope (BAT) catalog~\cite{Gehrels_2004} for energies below 250~keV and the INTEGRAL catalog~\cite{Bouchet_2008} for energies above 250~keV.
The Swift-BAT 105-month catalog includes 1632 sources with significances greater than 4.8$\sigma$ in the 14--195~keV band~\cite{Oh_2018}, while the INTEGRAL catalog lists 10 sources above 4.0$\sigma$ in the 200--600~keV band~\cite{Bouchet_2008}.
The resulting background light curve is shown as the green line in second panel of Fig.~\ref{fig:lightcurve}. 
\revise{
The normalization coefficient $A^{E'}$ was found to be 0.62, 0.80, 0.88, and 0.93 in the 150--250~keV, 250--350~keV, 350--450~keV,  and 450--600~keV bands, respectively.
}

\begin{figure*}
\begin{center}
\includegraphics[width=17cm]{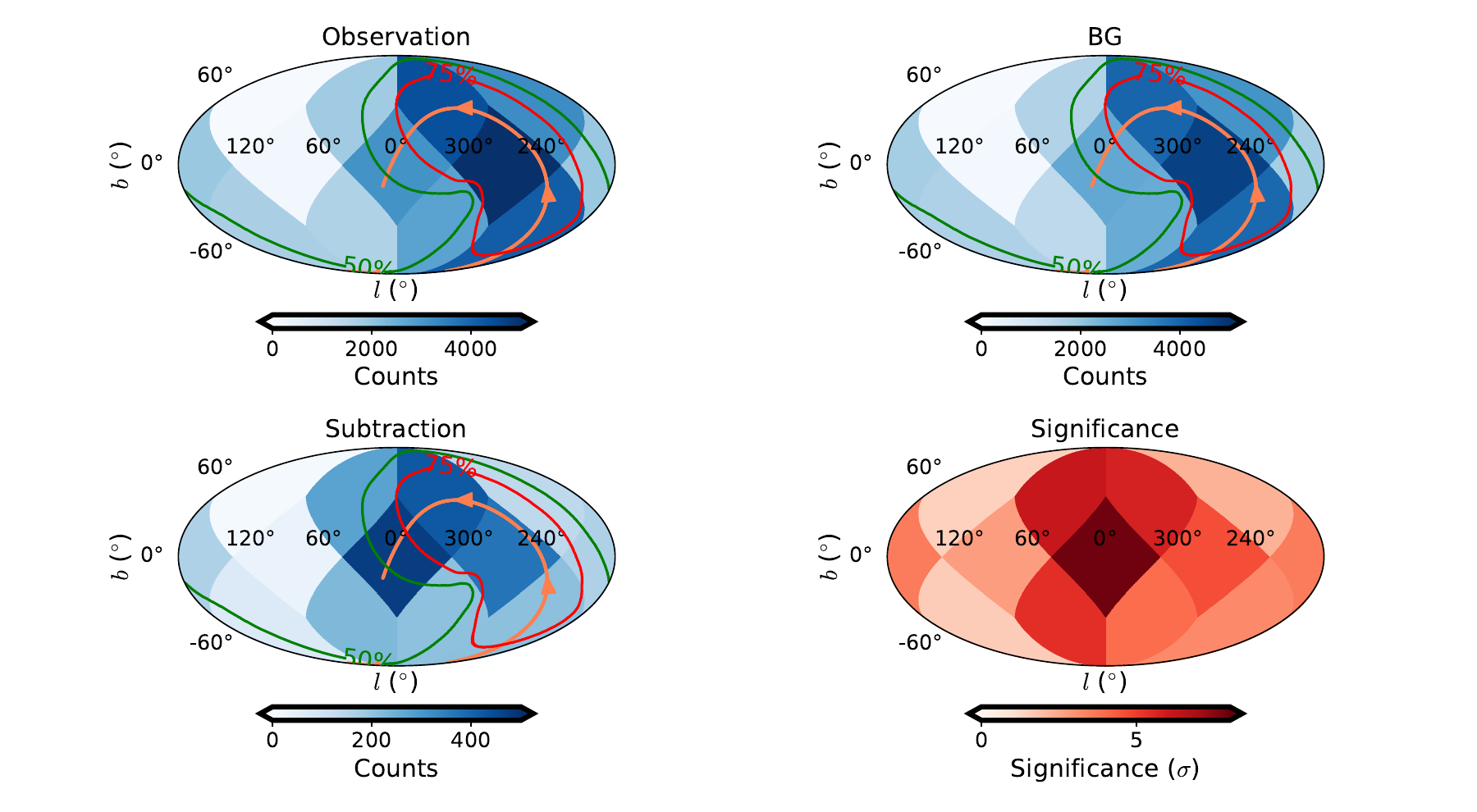}
\caption{\label{fig:skymap}
The top-left, top-right, and bottom-left panels show the observed skymap, the estimated background skymap, and the background-subtracted skymap, respectively, all displayed in Galactic coordinates.
The orange line traces the trajectory of the zenith direction of the ETCC during the observation. The red and green contours represent the 75\% and 50\% exposure levels at 511~keV, respectively. The bottom-right panel presents the significance map.}
\end{center}
\end{figure*}

\revise{In balloon-borne sub-MeV gamma-ray observations, cosmic gamma rays undergo attenuation and scattering in the atmosphere.
Atmospheric scattering enhances the detected flux in the low-energy region, and a fraction of the observed low-energy photons originates from this scattered component, thereby modifying the detected spectral shape.
To account for this atmospheric response, we simulated gamma-ray transport through the atmosphere at balloon altitude using Geant4.
This framework, which explicitly includes both the attenuation and scattering components, was developed in our previous study~\cite{Takada_2011}, and its consistency with the COSI atmospheric response model has been validated by the COSI collaboration~\cite{Karwin_2024}.}



Fig.~\ref{fig:skymap} presents the skymaps generated from the data collected during the stable altitude period between 2018-04-07T12:00 ACST and 2018-04-08T06:45 ACST, when the balloon maintained an altitude of approximately 39.6~km.
The top-left panel shows the observed skymap.
The top-right panel displays the background model skymap, and the bottom-left panel shows the background-subtracted skymap.
The orange line represents the trajectory of the ETCC zenith direction across the sky.
The red and green contours indicate the 75\% and 50\% exposure levels at 511~keV, respectively. The bottom-right panel displays the significance map, which includes the fitting uncertainty associated with the normalization coefficient $A^{E'}$.
The central pixel corresponding to the Galactic center shows a significant excess with a significance of 7.9$\sigma$. Additionally, the rightmost pixel along the Galactic plane, which includes the Crab Nebula, exhibits a significance of 3.6$\sigma$, as does the leftmost, consistent with our previous result~\cite{2022Takada}.


The diffuse source distribution $S^{E}(l,b)$ and the point sources $P^{E}(l,b)$ can, in principle, be determined by unfolding Eq.~(\ref{eq:conv}). However, due to the limited angular resolution, our data do not allow for a reliable separation of individual point sources.
To address this, we adopted external source catalogs, namely the Swift-BAT catalog for the energy range 150--250~keV and the INTEGRAL source catalog for 250--600~keV.
The low event counts preclude a fully model-independent reconstruction of the diffuse component $S^{E}(l,b)$.
Therefore, in this analysis, we tested three simplified emission models: ($\mathrm{i}$) a single point-like source at the Galactic center, ($\mathrm{ii}$) a multi-component model, and ($\mathrm{iii}$) a symmetric 2-dimensional (2D) Gaussian model.
The multi-component model includes a narrow bulge, a broad bulge, a low surface-brightness disk, and a central point source, with relative intensities and spatial extents fixed to the values reported in Ref.~\cite{Siegert_2016}.  The symmetric 2D Gaussian model is centered at $(l,b)=(0^{\circ},0^{\circ})$, with a common longitudinal and latitudinal width $\sigma_{\mathrm{sym}}=\sigma_{l}=\sigma_{b}$. 
In both the single point-like source and multi-component models, the total photon flux in each energy bin was treated as a free parameter, while in the symmetric 2D Gaussian model both the total photon flux and $\sigma_{\mathrm{sym}}$ were treated as free parameters. 

\section{Results}
We evaluated the goodness of fit for each emission model using the chi-square statistic. 
The resulting chi-square values (with corresponding $p$-values) were 43 (0.50) for the single point-like source model and 40 (0.63) for the multi-component model, indicating that both models are statistically consistent with the data.
For the symmetric 2D Gaussian model, the best-fit width is $\sigma_{\mathrm{sym}} = (26~\pm~12)^\circ$ corresponding to a full width at half maximum (FWHM) of $(61~\pm~28)^\circ$, which is approximately two times larger than the extent measured by COSI~\cite{Siegert_2020}.
On the other hand, an $F$-test comparing the 2D Gaussian model with the single point-like source model yields an $F$-statistic of 0.98 and a $p$-value of 0.32, indicating that the additional component of extent is not statistically significant. The fitting results are summarized in Table~\ref{tab:model}.


\begin{table*}
\caption{\label{tab:model}
Summary of emission model fitting results. Fluxes are in units of $10^{-2}\mathrm{~photons~cm^{-2}s^{-1}}$. Upper limits are given as a 95\% confidence level.
}
\begin{ruledtabular}
\begin{tabular}{lcccccc}
\textrm{Model}&
\textrm{150--250~keV}&
\textrm{250--350~keV}&
\textrm{350--450~keV}&
\textrm{450--600~keV}&
\textrm{Total} &
\textrm{ $\chi^{2}/\mathrm{d.o.f.}$}\\
\colrule
Single point-like source model & 2.6~$\pm$~0.8 & 1.4~$\pm$~0.4 & $<$0.8 & 1.0~$\pm$~0.5 & 5.0~$\pm$~1.0 & 43/44\\
Multi-component model          & 2.9~$\pm$~0.8 & 1.8~$\pm$~0.6 & $<$1.0 & 1.2~$\pm$~0.6   & 5.8~$\pm$~1.2 & 40/44\\
2D Gaussian model              & 3.0~$\pm$~0.8 & 1.9~$\pm$~0.6 & $<$1.0 & 1.3~$\pm$~0.6   & 5.8~$\pm$~1.3 & 39/43\\
\end{tabular}
\end{ruledtabular}
\end{table*}


The calculated light curve using the best-fit parameters from the multi-component model is shown by the red line in the third panel of Fig.~\ref{fig:lightcurve}, and shows good agreement with the increase associated with the Galactic center entering the FOV. The bottom panel displays the residuals, defined as the difference between the observed count rate and the total model including point sources, the multi-component model, and background. The fit accurately reproduces the observed light curve across the entire time window.
The relative contributions of point sources, the Galactic diffuse emission of the multi-component model, and background within the Galactic center bin are summarized in Table~
\ref{tab:proportion}.


\begin{table*}
\caption{\label{tab:proportion}
Composition of detected events and corresponding signal-to-noise ratios (\%) for each energy band within the Galactic center bin.
}
\begin{ruledtabular}
\begin{tabular}{lcccc}
\textrm{Model}&
\textrm{150--250~keV}&
\textrm{250--350~keV}&
\textrm{350--450~keV}&
\textrm{450--600~keV}\\
\colrule
Point sources        & 12 & 2.3 & 1.3 & 0.6 \\
Galactic diffuse emission & 12 & 13  & 4.0 & 7.5 \\
Background           & 75 & 85  & 95  & 92  \\
S/N                  & 32 & 18  & 5.6 & 8.8 \\
\end{tabular}
\end{ruledtabular}
\end{table*}

\revise{We calculated the gamma-ray intensity from the flux obtained through the multi-component spectral fit by accounting for the solid angle of the observed sky region.
In this analysis, the Galactic-center bin, defined as the region $|l|<33^{\circ}$ and $|b|<33^{\circ}$, was adopted as the normalization region, following the analysis binning scheme. }

\begin{figure}[h]
\includegraphics[width=8.5cm]{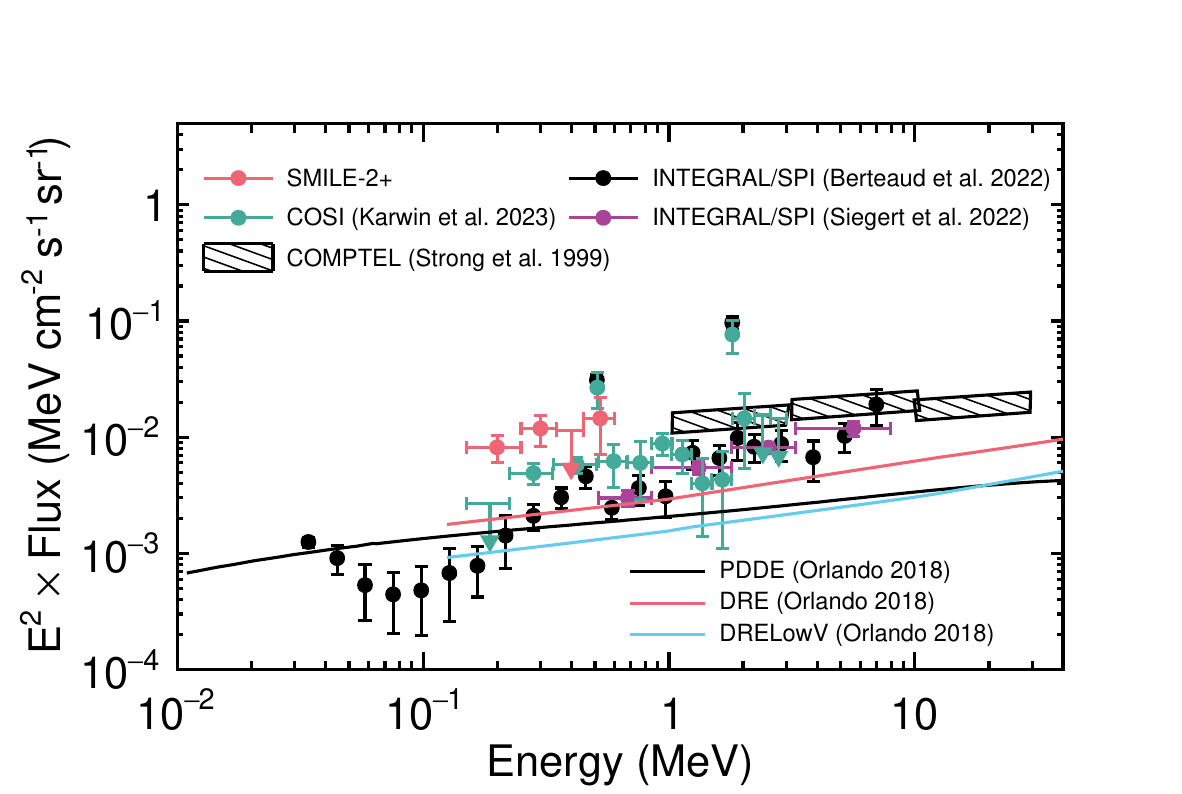}
\caption{\label{fig:spec} 
Measured gamma-ray intensity in the Galactic center region obtained with the ETCC, shown as a function of energy (red points with error bars), compared with previous measurements from INTEGRAL/SPI~\cite{Berteaud_2022}, and COSI~\cite{Karwin_2023}, and COMPTEL~\cite{Strong_1999}. Upper limits are shown at the 95\% confidence level.
The ETCC data correspond to the region $|l|<33^{\circ}$ and $|b|<33^{\circ}$, whereas the COSI and INTEGRAL measurements cover larger regions of $|l|<65^{\circ}, |b|<45^{\circ}$ and $|l|<47.5^{\circ}, |b|<47.5^{\circ}$, respectively. 
The black box and purple points with error bars indicate higher-energy measurements from COMPTEL and INTEGRAL/SPI~\cite{Siegert_2022}.
The solid black, red, and cyan curves represent inverse Compton emission models based on the PDDE, DRE, and DRELowV cosmic-ray propagation scenarios, for the region $|l|<33^{\circ}$ and $|b|<33^{\circ}$~\cite{Orland_2017}.}
\end{figure}

\revise{Fig.~\ref{fig:spec} shows the resulting gamma-ray intensity measured in the Galactic center bin, along with published values from other experiments. 
We note that for diffuse Galactic emission, the derived intensity depends on the adopted integration region and diffuse source distribution, and this comparison should therefore be regarded as approximate.
The sky coverage used in the COSI~\cite{Karwin_2023} and INTEGRAL/SPI measurements~\cite{Berteaud_2022} corresponds to $|l|<65^{\circ}$, $|b|<45^{\circ}$ and $|l|<47.5^{\circ}$, $|b|<47.5^{\circ}$, respectively. Hence, those intensity values in Fig.~\ref{fig:spec} are scaled by the different solid angles used in our analysis. 
The black, red, and cyan curves represent inverse Compton emission models calculated for the region $|l|<33^{\circ}$ and $|b|<33^{\circ}$, corresponding to the PDDE, DRE, and DRELowV baseline cosmic-ray propagation scenarios, respectively~\cite{Orland_2017}.
The spatial distributions for these models are available and the corresponding model fluxes can be integrated over the sky region covered by our field of view ($|l|<33^{\circ}$ and $|b|<33^{\circ}$). The model intensities shown here were therefore scaled to the same sky region used in this work, enabling a direct comparison with our measurement.}

In Sec.~\ref{sec:methods}, we used the Swift-BAT and INTEGRAL catalogs to subtract point source contributions from our data. In contrast, the INTEGRAL/SPI analysis subtracts point-source contributions using its own source catalog, whereas COSI did not perform any point-source subtraction in its analysis.


To calculate the positronium-related flux, we derived the total photon flux by integrating the emission map over the 150--600~keV band, giving $(5.8 \pm 1.2) \times 10^{-2}~\mathrm{photons~cm^{-2}~s^{-1}}$ in the multi-component model.
To isolate the positronium contribution, we estimated the inverse Compton flux obtaining $(2.6 \pm 0.7)~\times~10^{-2}~\mathrm{photons~cm^{-2}~s^{-1}}$, with the systematic uncertainty arising from differences between the DRE and DRELowV models.
\revise{After subtracting this component, the positronium-related flux is estimated to be $(3.2 \pm 1.4) \times 10^{-2}~\mathrm{photons~cm^{-2}~s^{-1}}$, which is consistent with the INTEGRAL value of $(1.4 \pm 0.3) \times 10^{-2}~\mathrm{photons~cm^{-2}~s^{-1}}$~\cite{Siegert_2016}, within 1$\sigma$.}

\section{Discussion}
A rough independent validation of the Galactic-center flux is obtained directly from the light curve data.
During the background pointing interval, the measured count rate can be well explained by the known extragalactic and atmospheric gamma-ray intensities~\cite{Ling}. The excess count rate observed in the light curve when the ETCC pointed toward the Galactic center provides a response-independent estimate of the Galactic center flux using the known extragalactic and atmospheric gamma-ray intensities.
In the 250--350~keV band, this estimate agrees within a factor of about 1.7 with the result obtained using the single  point-like source model in the imaging spectroscopy, indicating that our response model in the imaging analysis is not subject to severe biases.

In interpreting the positronium-related flux, we must account for the associated systematic uncertainties.
The intrinsic detector response was validated through ground calibration measurements using monochromatic sources, 
which showed that the largest deviation between the data and simulation is at the 12\% level at 356~keV. 
The atmospheric response was evaluated using a Monte Carlo simulation~\cite{Takada_2011}, 
which is consistent with that used in recent COSI analyses~\cite{Karwin_2024}. 
Assuming an $E^{-2}$ power-law spectrum, the balloon altitude fluctuation of $\pm 1$~km changes the predicted flux by at most $\pm 3.3$\% across zenith angles up to 60$^\circ$.

In our analysis, the limited energy resolution and statistics of the present measurement means that the diffuse continuum cannot be separated from the data alone, and model-dependent estimates must be adopted. Although we incorporate the spread among the three IC models as a systematic uncertainty, none of these models reproduce all the existing observations from INTEGRAL/SPI and COMPTEL simultaneously, and the true uncertainty may exceed the range spanned by these models. This modeling uncertainty is the primary limitation in achieving a fully self-consistent determination of the positronium-related flux in this work.

From the relative flux of the ortho-positronium (o-Ps) continuum and the 511~keV line, denoted by $I_{3\gamma}/I_{2\gamma}$, we evaluated the positronium fraction $f_\mathrm{{Ps}}$~\cite{Prantzos2011}. Here, $I_{3\gamma}$ and $I_{2\gamma}$ were calculated from the total fluxes in the 150--450 keV ($F_{150-450}$) and 450--600 keV ($F_{450-600}$) energy bins, respectively, with the latter bin capturing 99\% of the 511~keV gamma rays. 
Because the 450--600~keV bin is contaminated by the o-Ps continuum gamma rays due to the energy resolution, we calculated the flux ratio $F_{150-450}/F_{450-600}$ as a function of the positronium fraction using Monte Carlo simulations. 
\revise{The simulated value of $F_{150-450}/F_{450-600}$ was 1.6 for $f_\mathrm{{Ps}} = 1$, whereas the experimental value was $2.8~\pm~2.5$. 
The experimental value is statistically consistent with the simulated value within the uncertainty.}
\revise{The observed flux in the lowest-energy bin, 150--250~keV, has a strong impact on the value of $F_{150-450}/F_{450-600}$, while the continuum contribution is expected to be relatively small in this bin. 
On the other hand, the flux in the 350–450~keV bin, which should contain a larger fraction of the o-Ps continuum photons, yielded only an upper limit.
To assess the robustness of the result, we also evaluated the flux ratio using the 250--450~keV and 450--600~keV bands. The resulting experimental value, $F_{250-450}/F_{450-600} = 0.93 \pm 2.3$ is again consistent with the simulated value of 1.2 for $f_\mathrm{{Ps}} = 1$ within the large statistical uncertainty.
The current data provide only a limited constraint on the positronium fraction.
In particular, the large uncertainty reflects the limited photon statistics and the insufficient separation of the o-Ps continuum and 511~keV line components with the present energy resolution. }


\revise{
To quantitatively compare the background suppression and signal extraction performance, we evaluate the signal-to-noise ratio (S/N): the ratio of signal events to background events.
While the INTEGRAL/SPI experiment achieved a high statistical significance through long-term observations over more than a decade, it reported a signal fraction below 1\% due to a dominant instrumental background~\cite{Jean_2003}.
Despite the lack of the excellent energy resolution available to INTEGRAL/SPI, our measurement achieved a significantly higher S/N of 8.8\% at 450--600~keV.
This improvement can be primarily attributed to the ETCC’s superior background rejection, enabled by a full event-by-event kinematic consistency check~\cite{2023Ikeda}.}
Future upgrades to the angular resolution of recoil electron tracking are expected to further enhance the S/N. Simulations suggest that such improvements could increase the S/N to approximately 32\%.
For narrow spectral features such as the 511~keV positron annihilation line, the current energy resolution of 14\% limits the spectral sensitivity. Replacing the current scintillator with a semiconductor detector such as a CdZnTe detector~\cite{Stahle_1996} could dramatically improve the resolution and enhance line sensitivity, allowing the ETCC to perform high-precision spectroscopy of line emissions even from balloon platforms, opening the path to sensitive MeV-band Galactic surveys.


\section{Conclusion}
\revise{
We reported the detection of cosmic sub-MeV gamma rays from the Galactic center region using an ETCC aboard the SMILE-2+ balloon experiment.
Analyzing one-day flight data in the 150--600~keV range, we achieved a detection significance of 7.9$\sigma$ with a high S/N of 8.8\% in the 450--600~keV band, demonstrating the excellent background rejection capability of the ETCC.
The measured intensity and spatial distribution were tested against three emission models: a single point-like source, a multi-component structure, and a symmetric 2D Gaussian model. All the models were found to be statistically consistent with the observed data. 
The positronium-related flux for the multi-component model was $(3.2~\pm~1.4)~\times~10^{-2}$~photons~cm$^{-2}$~s$^{-1}$, which is consistent with the INTEGRAL result within the uncertainty.
These results validate the feasibility of precision measurements of diffuse Galactic gamma-ray emissions using the ETCC approach.
Future improvements in the angular resolution of electron track reconstruction are expected to further enhance the background suppression.
This result demonstrates, for the first time, that linear analysis using a rigorous PSF based on optical principles can be applied to MeV gamma-ray astronomy, as is standard practice at other wavelengths.
}


\begin{acknowledgments}
This study was supported by the Japan Society for the Promotion of Science (JSPS) KAKENHI Grant-in-Aids for Scientific Research (Grant Numbers 21224005, 20244026, 16H02185, 15K17608, 23654067, 25610042, 16K13785, 20K20428, 16J08498, 18J20107, 19J11323, 22J00064, 22KJ1766, and 24K00643), a Grant-in-Aid from the Global COE program “Next Generation Physics, Spun from Universality and Emergence” from the Ministry of Education, Culture, Sports, Science and Technology (MEXT) of Japan, and the joint research program of the Institute for Cosmic Ray Research (ICRR), The University of Tokyo. 
The balloon-borne experiment was conducted by researchers at Scientific Ballooning (DAIKIKYU) Research and Operation Group, ISAS, JAXA.
Some of the electronics development was supported by KEK-DTP and Open-It Consortium.
Furthermore, we would like to thank Naomi Tsuji for her insightful contributions.
The authors thank FORTE Science Communications (https://www.forte-science.co.jp/) for English language editing.
\end{acknowledgments}

%
%



\bibliography{apssamp}

\end{document}